\documentclass[twocolumn,showpacs]{revtex4}
\usepackage{amsmath}
\usepackage{graphicx}
\begin{document}

\title{Hysteresis in the de Haas-van Alphen Effect}
\author{R.B.G.~Kramer$^{a}$, V.S.~Egorov$^{a,b}$, A.G.M.~Jansen$^{c}$, and W.~Joss$^{a}$}
\affiliation{$^{a}$Grenoble High Magnetic Field Laboratory,
Max-Planck-Institut f\"{u}r Festk\"{o}rperforschung, B.P. 166X, F-38042 Grenoble Cedex 9, France \\
$^{b}$Russian Research Center "Kurchatov Institute", 123182 Moscow, Russia \\
$^{c}$Service de Physique Statistique, Magn\'{e}tisme, et
Supraconductivit\'{e}, D\'{e}partement de Recherche Fondamentale sur
la Mati\`{e}re Condens\'{e}e, CEA-Grenoble, F-38054 Grenoble Cedex
9, France}
\begin{abstract}
A hysteresis loop is observed for the first time in the de Haas-van
Alphen (dHvA) effect of beryllium at low temperatures and quantizing
magnetic field applied parallel to the hexagonal axis of the single
crystal. The irreversible behavior of the magnetization occurs at
the paramagnetic part of the dHvA period in conditions of Condon
domain formation arising by strong enough dHvA amplitude. The
resulting extremely nonlinear response to a \textit{very} small
modulation field offers the possibility to find in a simple way the
Condon domain phase diagram. From a harmonic analysis, the shape and
size of the hysteresis loop is constructed.
\end{abstract}
\date{\today}
\maketitle

It is well known that the irreversibility of the magnetization
process by domain wall motion is due to energy barriers arising from
a variety of defects inside magnetic material~\cite{SCHLENKER}. The
resulting hysteresis effects in usual magnetic substances, i.e. in
substances where the atomic magnetic moments, due largely to the
electron spin, are the reason of magnetism, have been investigated
in detail in the past. However, in an applied magnetic field, the
orbital motion of free electrons in metals leads also to
magnetization. For Landau quantization~\cite{LANDAU} of the
electronic system, the oscillating magnetization known as the de
Haas-van Alphen (dHvA) effect has been extensively studied in single
crystals at low temperatures~\cite{SHOENBERG}. There is no
experimental data reporting hysteresis in the dHvA effect apart from
the trivial case caused by eddy currents when the applied magnetic
field is varied fast enough~\cite{KNECHT}.

If the dHvA amplitude becomes very large and comparable to the
oscillation period, i.e. $\mu_0\partial M/\partial B>1$ for
magnetization $M$ and magnetic induction $B$, a thermodynamic
instability arises at the paramagnetic ($\chi=\partial M/\partial
H>0$ for susceptibility $\chi$ and applied magnetic field $H$) part
of every dHvA period, according to the Pippard-Shoenberg concept of
high magnetic interaction~\cite{SHOENBERG,PIPPARD}. For long samples
with the applied field parallel to the long axis (demagnetization
factor $n\approx0$) the instability is avoided by a jump of the
magnetic induction $B$ between the stable states $B_1$ and $B_2$ at
a critical field $H_c$. For plate-like samples perpendicular to
$\mathbf{H}$ ($n\approx1$) $B=\mu_0 H$ is required so that the
induction can not jump. In this case Condon domains arise with
neighboring regions of respective inductions $B_1$ and $B_2$ in the
applied field interval $B_1<\mu_0H<B_2$~\cite{CONDON}. For samples
with $0<n<1$ this interval decreases proportionally to $n$. For
samples of arbitrary shape there is an inhomogeneous state - Condon
domain state (CDS) with the same phases $B_1$ and $B_2$ .

After the discovery of Condon domains~\cite{CONDON,CONWAL}
hysteresis in the dHvA effect needs to be considered again. Indeed,
the CDS consists of two phases of different induction values with a
magnetization current in the domain walls. This needs usually extra
energy. It was shown that the transition from the homogeneous state
to the CDS is of first order~\cite{BLA95}. At this phase transition
one could expect, in principle, all phenomena like irreversibility,
supercooling and hysteresis that exist at first order phase
transitions, e.g. the liquid - gas transition. Naturally, Condon
discussed these problems in his first paper on domains~\cite{CONDON}
concluding that neither supercooling nor hysteresis had been
observed in all reported data. Since then, the above phenomena were
discussed in several papers~\cite{SMITH,GORDON}. Despite Condon
domains themselves have been observed and investigated
experimentally~\cite{VOLSKIJ,SOLT,SOLTEGOROV,LYKOV}, there is
nevertheless up to now no experimental data proving hysteresis.

In this Letter we present the first experimental observation of
hysteresis in beryllium single crystals. The hysteresis itself turns
out to be very small. Therefore, several methods were used to prove
the existence of hysteresis. First the hysteresis effect is measured
directly by Hall probes in DC fields, then a standard AC method is
used with various modulation levels, frequencies, and magnetic field
ramp rates. Finally, the DC hysteresis loop is reconstructed by
assembling several higher harmonics of the AC pickup voltage.
Moreover, it is shown that the Condon domain phase diagram can be
measured directly from the response to sufficiently small AC fields.
Finally, Plummer's~\cite{PLUMMER1964} strange and up to now only
incompletely understood data is explained.

Beryllium is to our knowledge the best metal to investigate
hysteresis effects related to Condon domain formation. First, due to
its cigar-like Fermi surface (small curvature at the maximal cross
sections) the dHvA amplitude is very high for
$\mathbf{H}\parallel[0001]$. Secondly, in this configuration two
rather close dHvA frequencies 970~T and 940~T coexist leading to a
beat in $M$ and $\chi$. Because of this beat there is the unique
possibility to change the dHvA amplitude and the critical parameter
$\chi$ by a factor three at constant $T$ by varying the magnetic
field only very little. Thus, experimental conditions can be
adjusted so that the transition to the inhomogeneous Condon domain
state occurs in a part of each beat period. The sample was cut from
the same piece that was used earlier for the preparation of the
plate-like sample in which Condon domains were first observed by
$\mu$SR~\cite{SOLT}. The results shown here were measured on a
rod-like sample of size $8\times 2\times 1$~mm$^3$ with the long
axis parallel to $[0001]$. The Dingle temperature was $T_D =2.0$~K.
The measurements were made either directly by a micro Hall probe
placed close to one end of the sample or by a compensated pickup
coil using low frequencies of about 21~Hz and small modulation level
($<6$~G). The experiments were carried out in a 10~T superconducting
coil with homogeneity better than 10~ppm in 1~cm$^3$.

\begin{figure}[tb]
     \begin{center}
   \leavevmode
       \includegraphics[width=\linewidth]{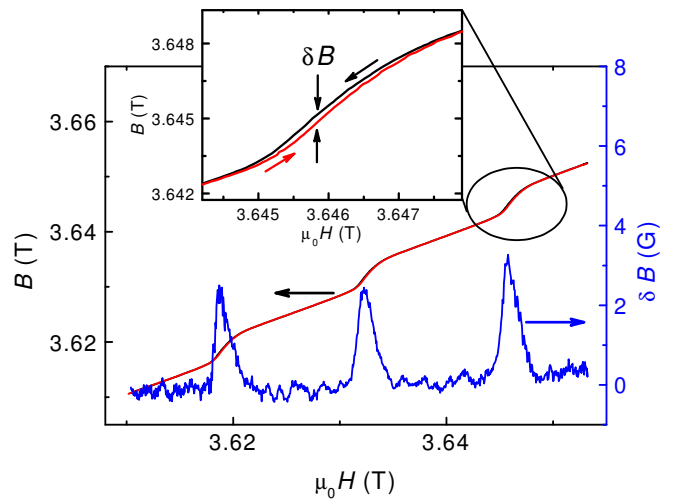}
       \caption{Hysteresis loop observed by Hall probes in the paramagnetic part of the dHvA oscillations of beryllium.
       $B(H)$ traces for an up and down sweep of the applied magnetic field around a dHvA antinode at
       3.6~T and $T=1.3$~K (scale on the left).
       The hysteresis loop is visible in the insert. The induction difference $\delta B$ between these curves
       shows the value of the hysteresis (scale on the right).}
    \label{FIG1}
    \end{center}
\end{figure}

Fig.~\ref{FIG1} shows Hall probe traces for an up and down sweep of
the applied magnetic field around a dHvA antinode at 3.6~T and at
$T=1.3$~K. The hysteresis loop is very small and only visible by a
zoom. A periodically arising induction difference $\delta B$ of
about 3~G between the up and down sweep is measured at the steeper
part, i.e. the paramagnetic part, of each dHvA period. The signal is
about ten times higher than the noise level. The applied field is
measured here by another Hall probe, placed sufficiently far from
the sample, as the superconducting solenoid has its own small
hysteresis when the current is swept. The hysteresis effect is
clearly observed in fig.~\ref{FIG1} using DC Hall probes but rather
under the most favorable conditions at 1.3~K around a maximum of a
magnetization beat. The DC Hall technique is not sensitive enough to
study hysteresis as a function of temperature and field.

A standard AC modulation method with a compensated coil system used
at \textit{very} low modulation level $h$ is much more sensitive to
detect nonreversible magnetization or hysteresis. This method is
used to determine exactly the point of its appearance. If the
modulation amplitude $h$ is much smaller than the oscillation period
$\Delta H$, the measured response corresponds in good approximation
to the derivative, i.e. the susceptibility $\chi(H)=\partial
M/\partial H$. If the modulation is further decreased, the result
should not change, the measurements should only be more precise.

\begin{figure}[tb]
     \begin{center}
   \leavevmode
       \includegraphics[width=\linewidth]{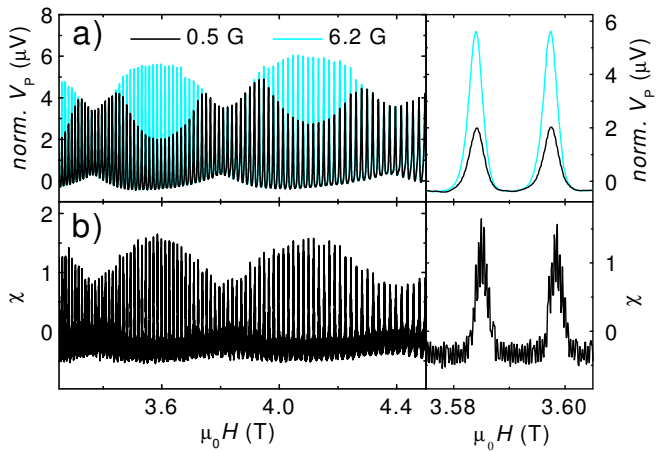}
       \caption{(a)~Pickup voltages divided by the modulation level
       for low and high modulation amplitude at 1.3~K. (b)~DC susceptibility
       derived from magnetization measurements with Hall probes without field modulation.
       Graphs to the right show respective zooms.}
    \label{FIG2}
    \end{center}
\end{figure}

Fig.~\ref{FIG2} shows a different behavior as a function of
modulation field amplitude $h$. All $\chi(H)$ curves of
Fig.~\ref{FIG2}(a) and (b) are measured at low temperature where in
the regions of each antinode hysteresis exists like it was shown on
Fig.~\ref{FIG1}. Fig.~\ref{FIG2}(a) shows the normalized pickup
voltage, i.e. response divided by the modulation level, measured by
the AC method with 6 and 0.5~G modulation amplitude. In both cases
$h\ll \Delta H$ so that one could expect identical curves. In fact,
both curves are completely congruent except at the regions near the
antinodes. Here, only the high modulation (6~G) level gives the
expected result, which is the well known dHvA oscillation beat of
beryllium. For comparison fig.~\ref{FIG2}(b) shows $\chi(H)$,
calculated from $B(H)$ curves measured with Hall probes like in
fig.~\ref{FIG1} without field modulation. For small modulation
amplitude deep "notches" in the dHvA oscillation envelope are
observed. The notches occur at magnetic fields where the dHvA
amplitude is big enough that the above described instability arises.
Instead of a further increase, after having crossed the critical
point, the dHvA amplitude decreases at the paramagnetic ($\chi>0$)
part of the dHvA period only. The diamagnetic part does not change.
One would expect that both $M(H)$ and $\chi(H)$ oscillate around
zero. This shows that the small modulation measurements can not
correspond to the real $\chi (H)$ wave form. The amplitude decrease
is absent at temperatures $T>3.4$~K and at the nodes of the dHvA
beat where the single crystal is in the homogeneous state.

\begin{figure}[tb]
     \begin{center}
   \leavevmode
       \includegraphics[width=0.9\linewidth]{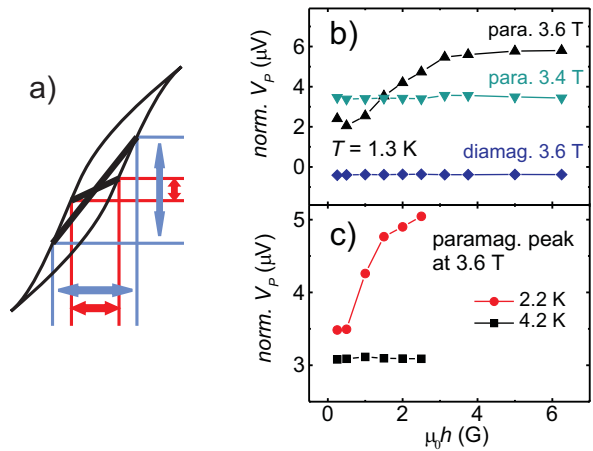}
       \caption{(a)~Schematic representation of the hysteresis loop
       showing that the response to an applied AC modulation field is nonlinear.
       (b)~Modulation level dependence of the normalized pickup
       voltage for characteristic magnetic fields at 1.3~K.
       (c)~The same dependence at the antinode at $T=2.2$~K and $T=4.2$~K.
       Nonlinearities arise at a critical temperature of 3.4~K.}
    \label{FIG3}
    \end{center}
\end{figure}

The decrease of the dHvA amplitude at low modulation level is
schematically shown on fig.~\ref{FIG3}(a). If the modulation is much
bigger than the width of the hysteresis $\delta H$ the AC response
corresponds in good agreement to the slope, i.e. $\chi(H)$. As soon
as the modulation becomes comparable to the hysteresis loop size the
response decreases~\cite{SCHLENKER}. The effect was measured for
several modulation levels at different temperatures and at different
magnetic fields with respect to the dHvA beat phase. The results are
shown on fig.~\ref{FIG3}(b) and~(c). The modulation level where the
dHvA amplitude decreases gives the hysteresis width $\delta H$ which
agrees with the above observed values of $\delta B$. No decrease is
observed at temperatures $T>3.4$~K and near the beat nodes.

This explanation is checked in a simple way. Usually the applied
magnetic field $H$ ramps much slower than the AC modulation field
$h$, i.e. the full magnetic field $H+h$ always oscillates around the
quasi static offset field $H$ and $B(H+h)$ makes a loop in presence
of hysteresis. If the ramp rates are changed in a way that
$\textrm{d}h/\textrm{d}t$ and $\textrm{d}H/\textrm{d}t$ are the
same, then the magnetic field sweeps only forward with small steps
in the direction of the ramp. Under this condition $B(H+h)$ never
makes a loop as we go always along the hysteresis loop boundary and
never inside. In this regime, which is usually not used, the lock-in
amplifier does not measure the correct amplitude. We observed
however in this regime the usual dHvA beat signal without notches.
\begin{figure}[tb]
     \begin{center}
   \leavevmode
       \includegraphics[width=0.9\linewidth]{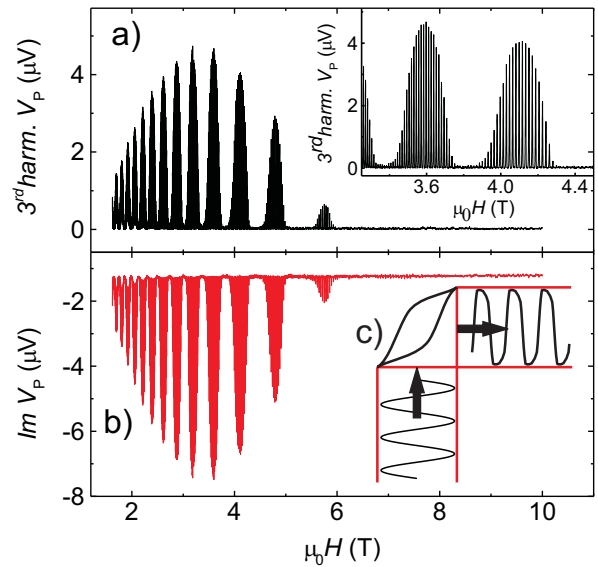}
       \caption{(a)~Third harmonic of the pickup voltage. The insert shows a
       zoom to the above discussed field range. (b)~Out of phase part of the
       pickup voltage. (both at 2.5~G modulation level and 1.3~K).
       The insert shows a schematic representation of the hysteresis showing
       that the response to a sinusoidal field modulation becomes window shaped and
       is slightly phase shifted with respect to the input.}
    \label{FIG4}
    \end{center}
\end{figure}

The notches in the envelope of the first harmonic in-phase AC
response are "compensated" by steeply rising higher harmonics at the
same critical point. At the same field a phase shift appears in the
pickup voltage. This means that the response to an AC modulation
becomes extremely nonlinear in the presence of hysteresis.
Fig.~\ref{FIG4} shows the third harmonic~(a) and the imaginary
part~(b) of the pickup voltage in a wide region of magnetic field at
1.3~K. The amplitude is big in both curves only around the beat
antinodes. In the regions of nodes the signal is about zero. The
insert of fig.~\ref{FIG4}(a) shows the $3^{\mathrm{rd}}$ harmonic in
the same field interval as in fig.~\ref{FIG2}. The comparison shows
that the signal appears and disappears with a threshold character at
the critical points of the transition to the CDS. The insert in
fig.~\ref{FIG4}(b) shows schematically that the response to an
initially sinusoidal modulation field becomes highly distorted in
the presence of a hysteresis loop. The response becomes more
rectangular shaped like a window-function. This function can be
composed, as it is well known, of odd harmonics of a sine. This
explains that the third harmonic content of the pickup voltage is
very high in the CDS. At magnetic fields without hysteresis the
$3^{\mathrm{rd}}$ harmonic is very small as the modulation level
$h=2.5$~G is much less than the dHvA period of about 130~G at 3.6~T.
\begin{figure}[tb]
     \begin{center}
   \leavevmode
       \includegraphics[width=0.85\linewidth]{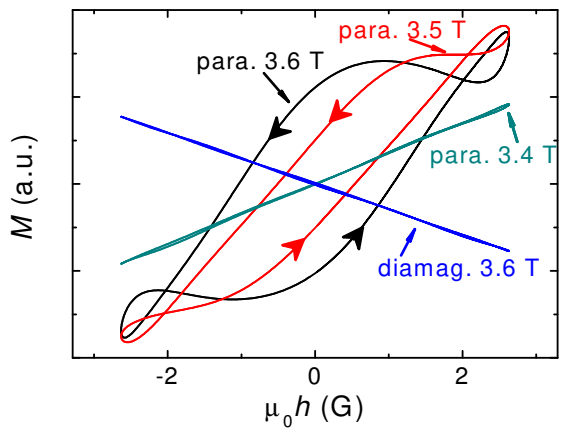}
       \caption{Hysteresis shape reconstruction for
       modulation amplitude 2.5~G for several characteristic magnetic fields at 1.3~K.}
    \label{FIG5}
    \end{center}
\end{figure}

The hysteresis shape and size is reconstructed in fig.~\ref{FIG5}
like e.g. in~\cite{Ruedt2004}. The response to the sinusoidal
modulation of 2.5~G amplitude is calculated by adding all in and out
of phase contributions up to the fifth harmonic. The same procedure
was applied at different positions along the dHvA oscillations. At
the diamagnetic part ($\chi<0$) of every dHvA period all harmonics
vanish and only the in phase response persists. Hence, a line with
negative slope is calculated. The same behavior is found for
$\chi>0$ around a node with a line with positive slope. Whereas at
the paramagnetic part in the region of the notches hysteresis arises
and its size is maximal at the dHvA beat antinode.

The observed threshold behavior of the third harmonic and of the out
of phase signal does not essentially change when the modulation
level is varied providing $h\ll \Delta H$. This behavior was
observed in the frequency range from 8~Hz up to about 200~Hz. Thus,
these measurements offer a simple way to determine the Condon domain
phase diagram.

Moreover, hysteresis may explain the data measured by
Plummer~\cite{PLUMMER1964} with an AC mutual inductance method at
low modulation level. First Plummer thought that a new dHvA
frequency was discovered. As the Fermi surface of beryllium was not
consistent with this frequency, eddy currents at the rather high
frequency of 100~Hz were invoked~\cite{PLUMMER1966,KNECHT}. A
comparison of Plummer's data with our results (see
Fig.~\ref{FIG2}(a), black curve) shows the similarity. The deep
notches in the dHvA oscillation envelope seem indeed to be the
result of a new frequency. We repeated the measurements at higher
frequencies and found the same behavior.

Hysteresis accompanies without doubt the appearance of the CDS
resulting from the Pippard-Shoenberg instability. For any sample
shape, not only in plate-like samples, in the CDS coexist two phases
$B_1$ and $B_2$ with domain walls between them. The wall motion and
pinning might, in principle, depend on the sample shape. Therefore,
a plate-like sample was measured in the same $T$ and $H$ range. We
found the same phase diagram for hysteresis formation in both
samples. The hysteresis size, however, needs to be investigated more
precisely as function of temperature, magnetic field and for various
$n$.

Hysteresis itself is a result of interaction or pinning of domain
walls with defects, impurities and the surface of the crystal.
Moreover, the question of domain wall motion can not be neglected in
many phenomena like e.g. acoustic wave propagation and absorption,
and helicon waves~\cite{VOLSKIJ}. Recently, the question of Condon
domain wall motion was considered in detail
theoretically~\cite{GORDONPRB}. Unfortunately, an idealized model of
a domain wall was used without taking into account the direct link
with lattice deformation~\cite{LYKOV} and the metastable behavior
due to hysteresis.

In conclusion we have shown that hysteresis occurs in the dHvA
effect under the conditions of the Condon domain state. The observed
hysteresis loop width is rather small, only a few Gauss. We have
shown that the out of phase part and the third harmonic of the
pickup voltage rise steeply when the magnetization becomes
irreversible. This threshold character offers a simple and robust
possibility to measure a Condon domain phase diagram.

We are grateful to I.D.~Vagner and I.~Sheikin for fruitful
discussions.


\begin{thebibliography}{99}

\bibitem{SCHLENKER} E.~du Tr\'{e}molet de Lacheisserie {\it et al.}, {\it Magnetism},
(Kluwer Academic Publishers, Norwell, 2002).

\bibitem{LANDAU} L.~Landau, Z. Phys. {\bf 64}, 629
(1930).

\bibitem{SHOENBERG} D.~Shoenberg, {\it Magnetic Oscillations in Metals},
(Cambridge University Press, Cambridge, 1984).

\bibitem{KNECHT} B.~Knecht {\it et al.}, J. Low Temp. Phys. {\bf 29}, 499
(1977).

\bibitem{PIPPARD} A.B.~Pippard, Proc. Roy. Soc. {\bf A272},
192 (1963).

\bibitem{CONDON} J.H.~Condon, Phys. Rev. {\bf 145}, 526 (1966).

\bibitem{CONWAL} J.H.~Condon and R.E. Walstedt, Phys. Rev. Lett. {\bf 21}, 612 (1968).

\bibitem{BLA95} Ya.M.~Blanter, M.I.~Kaganov, and D.V.~Posvyanskii, Sov. Phys.
Usp., {\bf 38}, 203 (1995).

\bibitem{SMITH} J.L.~Smith and J.C.~Lashley, J. Low Temp. Phys.
{\bf 135}, 161 (2004).

\bibitem{GORDON} A.~Gordon, I.D.~Vagner, and P.~Wyder, Adv. in Physics {\bf 52}, 385
(2003).

\bibitem{VOLSKIJ} V.I.~Bozhko and E.P.~Volskij, JETP Lett. {\bf 26}, 223 (1977).

\bibitem{SOLT} G.~Solt {\it et al.}, Phys. Rev. Lett. {\bf 76}, 2575 (1996).

\bibitem{SOLTEGOROV} G.~Solt and V.S.~Egorov, Physica B {\bf 318}, 231
(2002).

\bibitem{LYKOV} V.S.~Egorov and P.V.~Lykov, Sov. Phys. JETP, {\bf 94},
162 (2002).

\bibitem{PLUMMER1964} R.D.~Plummer and W.L.~Gordon, Phys. Rev. Lett. {\bf 13}, 432 (1964).

\bibitem{PLUMMER1966} R.D.~Plummer and W.L.~Gordon, Phys. Lett. {\bf 20}, 612 (1966).

\bibitem{Ruedt2004} C.~R\"udt {\it et al.}, Phys. Rev. B {\bf 69},
014419 (2004).

\bibitem{GORDONPRB} A. Gordon, N. Logoboy, and W. Joss,
Phys. Rev. B {\bf 69}, 174417 (2004)

\end{thebibliography}
\end{document}